\documentclass{rcdj}

\usepackage{hhline}
\makeatletter
\renewcommand*\l@paragraph{\@dottedtocline{4}{1.0em}{2.1em}}
\@addtoreset{paragraph}{section}
\makeatother

\begin{document}

\setcounter{page}{201}
\journal{REGULAR AND CHAOTIC DYNAMICS, V.\,7, \No2, 2002}
\title{THE ROLLING MOTION OF A BALL ON A SURFACE.\\
NEW INTEGRALS AND HIERARCHY OF DYNAMICS}
\runningtitle{THE ROLLING MOTION OF A BALL ON A SURFACE}
\runningauthor{A.\,V.\,BORISOV, I.\,S.\,MAMAEV, A.\,A.\,KILIN}
\authors{A.\,V.\,BORISOV}
{Department of Theoretical Mechanics\\
Moscow State University,
Vorob'ievy Gory\\
119899, Moscow, Russia\\
E-mail: borisov@rcd.ru}
\authors{I.\,S.\,MAMAEV}
{Laboratory of Dynamical Chaos and Nonlinearity\\
Udmurt State University, Universitetskaya, 1\\
426034, Izhevsk, Russia\\
E-mail: mamaev@rcd.ru}
\authors{A.\,A.\,KILIN}
{Laboratory of Dynamical Chaos and Nonlinearity\\
Udmurt State University, Universitetskaya, 1\\
426034, Izhevsk, Russia\\
E-mail: aka@rcd.ru} \amsmsc{37J60, 37J55} \received 10.02.2002.
\doi{10.1070/RD2002v007n02ABEH000205} \abstract{The paper is concerned
with the problem on rolling of a homogeneous ball on an arbitrary surface.
New cases when the problem is solved by quadratures are presented. The
paper also indicates a special case when an additional integral and
invariant measure exist. Using this case, we obtain a nonholonomic
generalization of the Jacobi problem for the inertial motion of a point on
an ellipsoid. For a ball rolling, it is also shown that on an arbitrary
cylinder in the gravity field the ball's motion is bounded and, on the
average, it does not move downwards. All the results of the paper
considerably expand the results obtained by E.\,Routh in XIX century.}

\maketitle

\section*{Contents}

\makeatletter
\@starttoc{toc}
\makeatother

\vspace{1cm}

\section{Introduction}

In this paper we consider the problem of sliding-free rolling of a
dynamically symmetrical ball (the central tensor of inertia is
spherical~${\bf I}=\mu{\bf E}$) on an arbitrary surface. As it was
indicated by E.\,Routh in his famous treatise~\cite{Raus}, if the surface
is a surface of revolution the problem is integrable even in the presence
of axisymmetric potential fields. Here we give a more complete analysis of
the Routh solution for the body of revolution, and present new integrals
for the case of ball rolling on non-symmetrical surfaces of the second
order.

\section{The equations for a ball moving on a surface}

In rigid body dynamics it is customary to introduce a body-fixed frame of
reference. However, while studying motion of a homogeneous ball, it is
more convenient to write the equations of motion with respect to a certain
fixed in space frame of reference. In such a frame a balance of linear
momentum and of angular momentum with respect to the ball's center of
mass, involving the reaction and the external forces, may be written as
\eq[eq1-1]{ m\dot \bv=\bN+\bF,\qquad ({\bf I}\bs\om)^{\ds.}=\ba\x
\bN+\bM_{\bF}. }

\wfig<bb=0 0 45.3mm 33.7mm>[15]{ris3.eps}[The rolling of a ball on a
surface ($G$ is the center of mass, $Q$ is a point of contact with the
surface)]

\noindent The sliding-free condition reads (the contact point velocity
vanishes) \eq[eq1-2]{ \bv+{\bs\om}\x \ba=0. } Here~$m$ is the mass of the
ball, $\bv$ isthe velocity of its center of mass, $\bs\om$ is the angular
velocity, ${\bf I}=\mu {\bf E}$ is the (spherical) central tensor of
inertia, $\ba$ is the vector from the center of mass to the point of
contact, $R$ is the ball's radius, $\bN$ is the reaction at the contact
point (see Fig.~\ref{ris3.eps}), $\bF$ and~$\bM_F$ are the external force
and the moment of forces with respect to the point of contact
respectively.

Eliminating from these equations the reaction $\bN$ and making use of the
fact that the contact point velocities on the surface and on the ball
coincide, we obtain the system of six equations:
\begin{equation}
\label{eq1-3} \dot{\bM} = D\dot{\bs\gam}\x ({\bs\om}\x{\bs\gam})+\bM_F,
\quad \dot{\br}+R\dot{\bs\gamma}=\bs\omega\x R\bs\gamma.
\end{equation}
Here $D=mR^2$. These equations govern the behaviour of the vector of the
angular momentum (with respect to the contact point $\bM$), and of the
vector $\bs\gamma=-R^{-1}\ba$ normal to the surface (Fig.~\ref{ris3.eps}).
The vectors $\bs\omega$ and $\br$ (the position vector of the contact
point) are obtained from the relations
\begin{equation}
\label{eq1-4}
\bM=\mu \bs\omega+D\bs\gamma\x (\bs\omega\x\bs\gamma),\quad
\bs\gamma=\frac{\nabla F(\br)}{|\nabla F(\br)|},
\end{equation}
$F(\br)=0$ is the equation of the surface on which the ball rolls. The
last equation in~\eqref{eq1-4} defines the Gaussian map. Further on,
following Routh, we will use the explicit form of the surface on which
\emph{the ball's center of mass} is moving. This surface, defined by the
position vector $\br'=\br+R\bs\gamma$, is equidistant relative to the
surface on which the contact point is moving.

In the case of potential forces, the moment $\bM_F$ is expressed in terms
of the potential $U(\br')=U(\br+R\bs\gamma)$. This potential depends on
the position of the ball's center of mass as
follows~$\bM_F=R\bs\gamma\times \pt{U}{\br'}$.

\begin{rem}
In his treatise~\cite{Raus} Routh obtained the equations of motion of the
ball in semifixed axes and presented the cases when these equations can be
analytically solved. Actually, in the majority of the subsequent
publications~\cite{Appell,Neimark} Routh's results were just restated
without any essential expansion. It should be noted that Routh was
especially interested in the stability of particular solutions (e.\,g. a
ball rotating around the vertical axis at the top of a surface of
revolution). Here we are not giving the original form of Routh's
equations. Equations~(\ref{eq1-3}) are, in many respects, similar to the
equations describing an arbitrary body motion on a plane and a
sphere~\cite{Bormam-new}. This allows us to consider many problems (e.\,g.
those of integrability) from the single point of view.
\end{rem}

\paragraph{The integrals of motion.}
In the case of potential field with potential~$U(\br+R\bs\gamma)$,
equations~(\ref{eq1-3}) possess the integral of energy and the geometric
integral
\begin{equation}
\label{eq1-5}
H=\frac12(\bM,\bs\om)+U(\br+R\bs\gamma), \qq F_1=\bs\gam^2=1.
\end{equation}
In the case of an arbitrary surface $F(\br)=0$, apart from these
integrals, the system~\eqref{eq1-3} has neither measure, nor two
additional integrals, which are necessary for the integrability according
to the last multiplier theory (the Euler\f Jacobi theory). The system's
behavior is chaotic. As it is shown later, in some cases there exists a
measure and only one additional integral. In such a case chaos becomes
``weaker''. As it was noted by Routh, for a surface of revolution there
exist two additional integrals, the system is integrable, and its behavior
is regular. The reduced system becomes a Hamiltonian one after an
appropriate change of time.

\paragraph{Rolling on a surface of the second order.}
We are going now to revisit equations\eqref{eq1-3} for the case when the
ball's center of mass is moving on a surface of the second order. The
surface is defined by the equation
\begin{equation}
\label{eq1-6}
\bigl(\br+R\bs\gamma,{\bf B}^{-1}(\br+R\bs\gamma)\bigr)=1,\qquad
{\bf B}=\diag (b_1,b_2,b_3)
\end{equation}
(for an ellipsoid, $b_i>0$ and are equal to the squares of the main
semi-axes). Solving \eqref{eq1-6} for the position vector~$\br$ gives
\begin{equation}
\label{eq1-7} \br+R\bs\gamma= \frac{{\bf
B}\bs\gamma}{\sqrt{(\bs\gamma,{\bf B}\bs\gamma)}}.
\end{equation}
We obtain the equations of motion in terms of the variables
$\bM,\bs\gamma$:
\begin{equation}
\label{eq1-8}
\dot{\bM}=-\frac{D}{\mu+D}(\bM,\dot{\bs\gamma})\bs\gamma,\qquad
\dot{\bs\gamma}=\frac{R\sqrt{(\bs\gamma,{\bf B}\bs\gamma)}}
{\mu+D}\bs\gamma\x\bigl(\bs\gamma\x{\bf B}^{-1}(\bs\gamma\x\bM)\bigr)
\end{equation}

\paragraph{A ball on a rotating surface.}
Let us also consider the motion a ball on a surface rotating with constant
angular velocity~$\bs\Om$. One particular case of this problem (a plane
and a sphere) was also investigated by Routh~\cite{Raus}. By analogy with
the previous case, replacing the nonholonomic relation~\eqref{eq1-2} by
\begin{equation}
\label{eq1-9}
\bv+\bs\om\x \ba=\bs\Om\x \br,
\end{equation}
we get
\begin{equation}
\label{eq1-10}
\begin{gathered}
\dot {\bM}  = m\bigl(\dot {\ba}\x(\bs\om\x \ba)+\ba\x(\bs\Om\x \dot {\br})\bigr)+ \bM_F\\
\dot {\br}+R\dot{\bs\gamma}  = \bs\omega \x R\dot{\bs\gamma}+ \bs\Om\x
\br.
\end{gathered}
\end{equation}
Here~$\ba=-R\bs\gam$.\goodbreak

It should be noted that in the case of an arbitrary surface the system of
six equations~\eqref{eq1-10} is not closed because the position vector of
the point on the surface, $\br$, is not expressed in terms of $\bs\gamma$
only; one should also introduce the equation for the angle of rotation of
the surface around the fixed axis. Nevertheless, if the surface is
axisymmetric, and the symmetry axis coincides with the axis of rotation,
equations~\eqref{eq1-10} get closed. We assume this to be fulfilled in
further text.

Equations \eqref{eq1-2}, \eqref{eq1-10} are in many respects similar to
the equations, defining the rigid body rolling on a plane or a sphere,
which we thoroughly investigated in the paper~\cite{Bormam-new}. We shall
use this fact while transferring the corresponding results onto
systems~\eqref{eq1-2}, \eqref{eq1-10}.

\section{Ball's motion on a surface of revolution}

First of all, let us consider the cases of integrability of
equations~\eqref{eq1-3}, \eqref{eq1-10}, associated with the rotational
symmetry of the surface on which the ball rolls. Here we are using the
technique introduced in~\cite{Bormam-new} and concerned with the analysis
of a certain reduced system in terms of new variables~$K_1$, $K_2$, $K_3$,
$\gamma_3$. We shall also assume the surface to be rotating around the
axis of symmetry with constant angular velocity $\bs\Om=(0,0,\Om)$,
$\Omega\ne 0$. The particular case, $\Omega=0$, was discussed by Routh,
who obtained the majority of the following results (although he missed
some cases of equal interest).

The equation of a surface of revolution with respect to an immovable
reference frame may be written as
\begin{equation}
\label{eq1-11}
\begin{gathered}
r_1=(f(\gam_3)-R)\gam_1, \q r_2=(f(\gam_3)-R)\gam_2, \\
r_3=\int \Bigl(f(\gam_3)-\frac{1-\gamma_3^2}{\gamma_3}f'(\gamma_3)
\Bigr)d\gamma_3
-R\gamma_3,
\end{gathered}
\end{equation}
where $f(\gamma_3)$ is a certain function specifying the surface
parametrization. The parametrization \eqref{eq1-11} is to be so chosen as
to make the form of the reduced system as simple as possible.

In the case being considered equations \eqref{eq1-3}, \eqref{eq1-10} allow
an invariant measure with density
\begin{equation}
\label{eq1-12}
\rho = \bigl(f(\gamma_3)\bigr)^3g(\gamma_3), \quad \text{where}
\quad g(\gamma_3)=f(\gamma_3)-\frac{1-\gamma_3^2}{\gamma_3}f'(\gamma_3).
\end{equation}
Apart from the invariant measure, the equations also have a simple
symmetry field
\begin{equation}
\label{eq1-13}
\hat{\bv}=M_1\pt{}{M_2}-M_2\pt{}{M_1}+\gamma_1\pt{}{\gamma_2}-\gamma_2
\pt{}{\gamma_1},
\end{equation}
which is caused by the rotational symmetry.

In our book \cite{BorisovMamaev1} we make a frequent use of the fact that
to obtain the simplest form of the reduced system in the presence of
symmetries, one needs to choose the most relevant {\em integrals of the
field of symmetries} which, in this case, may be written as
\begin{equation}
\label{eq1-14}
K_1=(\bM,\bs\gamma)f(\gamma_3),\quad K_2=\mu\omega_3=
\frac{\mu M_3+D(\bM,\bs\gamma)\gamma_3}{\mu+D},\quad
K_3=\frac{M_1\gamma_2-M_2\gamma_1}{\sqrt{1-\gamma_3}},\quad \gamma_3.
\end{equation}
In terms of the chosen variables, the reduced system takes the form
\begin{equation}
\label{eq1-15}
\begin{gathered}
\dot\gamma_3=k K_3,\\
\dot K_1=kK_3\Bigl(\frac{f'}{\gamma_3}K_2+\Bigl(1-\frac{f}{R}\Bigr)
g\mu\Omega\Bigr),\\
\dot K_2=kK_3\frac{D}{\mu +D}\Bigl(\frac{1}{f}
K_1-\gamma_3\Bigl(1-\frac{g}{R}\Bigr)\mu\Omega\Bigr),\\
\dot K_3=-k\frac{(\mu+D)g}{\mu^2(1-\gamma_3^2)f^2}
\Bigl(\frac{\gamma_3K_1-fK_2}{f(1-\gamma_3^2)}\bigl(
(\mu+D\gamma_3^2)K_1-\gamma_3 f(\mu+D)K_2\bigr)+\\
+\Bigl(1-\frac{g}{R}\Bigr)\bigl((\mu+2D\gamma_3^2)K_1-\gamma_3f
(\mu+2D)K_2\bigr)\Bigr)\mu\Omega,
\end{gathered}
\end{equation}
where $k=\frac{R\sqrt{1-\gamma_3^2}}{(\mu+D)g(\gamma_3)}$. It is easy to
show that this system of equations possesses an invariant measure with
density $\rho=k^{-1}$. The system~\eqref{eq1-15} can be explicitly
integrated in the following way.

Let us divide the second and the third equation of system~\eqref{eq1-15}
by~$\dot\gamma_3$. Then we obtain the nonautonomous system of two linear
equations with the independent variable~$\gamma_3$
\begin{equation}
\label{eq1-16}
\frac{dK_1}{d\gamma_3}=\frac{f'}{\gamma_3}K_2+
\Bigl(1-\frac{f}{R}\Bigr)g\mu\Omega,\quad
\frac{dK_2}{d\gamma_3}=\frac{D}{\mu+D}
\Bigl(\frac{1}{f}K_1-\gamma_3\Bigl(1-\frac{g}{R}\Bigr)\mu\Omega\Bigr).
\end{equation}
This system of linear equations always possesses two integrals which are
linear in~$K_1$, $K_2$. The coefficients in the integrals are functions
of~$\gamma_3$, and in the general case cannot be obtained in the explicit
(algebraic) form. Having divided the last equation of the
system~\eqref{eq1-15} by~$\gamma_3$ and substituting into this equation
the known solution of the system~\eqref{eq1-16}, we obtain the explicit
quadrature for~$K_3(\gamma_3)$. Using the first equation
from~\eqref{eq1-15} we can obtain the expression for~$\gamma_3(t)$.

In the case $\Omega=0$, the system~\eqref{eq1-15} has the integral of
energy
\begin{equation}
\label{eq1-18} H=(\bM,\bs\omega)=\frac12\biggl(\frac{K_1^2}{\Lambda f^2}+
\frac{(\mu+D)(\gamma_3K_1-fK_2)^2}{\mu^2f^2(1-\gamma_3^2)}
+\frac{K_3^2}{\mu+D}\biggr).
\end{equation}
The quadrature for $\gamma_3(t)$ may be obtained upon substitution of
$K_3=k^{-1}\dot\gamma_3$ into ~\eqref{eq1-18}.

For the reduced system~\eqref{eq1-15} the following theorem is valid:

\begin{teo}
For $\Omega=0$ by the change of time $k\,dt=d\tau$, the
system~\eqref{eq1-15} can be represented in the Hamiltonian form
\begin{equation}
\label{eq3-8} \frac{dx_i}{d\tau}=\{x_i,H\},\quad
\bx=(\gamma_3,K_1,K_2,K_3).
\end{equation}
The bracket is degenerate and specified by the relations
\begin{equation}
\label{eq3-9}
\{\gamma_3,K_3\}=\mu+D,\quad
\{K_1,K_3\}=(\mu+D)\frac{f'}{\gamma_3}K_2,\quad \{K_2,K_3\}=\frac{D}{f}K_1
\end{equation}
{\rm(}for the other pairs of variables the brackets are zero\/{\rm)}.
\end{teo}

\proof*() is a straightforward exercise consisting in derivation
of~~\eqref{eq1-15} ($\Omega =0$) from~\eqref{eq3-8} and verification of
the Jacobi identity for the bracket~\eqref{eq3-9}.\qed

One can assert that the equations for the linear integrals of the
system~\eqref{eq1-16} exactly coincide with the equations for the Casimir
function of the bracket~\eqref{eq3-9}.

It should also be noted that for $\Omega =0$ the system~\eqref{eq1-15} has
``skew-symmetrical notation'', similar to the Hamiltonian
form~\eqref{eq3-8}, \eqref{eq3-9}
\begin{equation}
\label{eq3-*1} \frac{dx_i}{d\tau}=J_{\lambda ij}(x)\pt{H}{x_j};\quad
J_{ij}=-J_{ji}.
\end{equation}
The matrix ${\bf J}_\lambda$ is of a somewhat more general form than that
of the corresponding bracket~\eqref{eq3-9}
\begin{equation}
\label{eq3-*2}
\begin{gathered}
{\bf J}_\lambda = \begin{pmatrix}
0 & 0 & 0 & \mu +D\\
0 & 0 & \lambda & (\mu+D)\frac{f'}{\gamma_3}K_2+\lambda u\\
0 & -\lambda & 0 & \frac {D}{f}K_1+\lambda v\\
-(\mu+D) & -(\mu+D)\frac{f'}{\gamma_3}K_2-\lambda u &
\frac{D}{f}K_1+\lambda v & 0
\end{pmatrix},\\
u=\frac{(\mu+D)^2(\gamma_3K_1-fK_2)}{\mu^2f(1-\gamma_3^2)K_3},\quad
v=\frac{(\mu+D)\bigl((\mu+D\gamma_3^2)K_1-f\gamma_3(\mu+D)K_2\bigr)}
{\mu^2f^2(1-\gamma_3^2)K_3}.
\end{gathered}
\end{equation}
Here $\lambda$ is an arbitrary function of $\gamma_3$, $K_i$. For
$\lambda=0$, one again gets bracket~\eqref{eq3-9}, and the tensor ${\bf
J}_\lambda$ satisfies the Jacobi identity. However, in the general case,
${\bf J}_\lambda$ does not satisfy the Jacobi identity (it is not of the
Poisson type). Nevertheless, if we assume
\begin{equation}
\label{eq3-*3}
\lambda=w(\gamma_3)K_3,
\end{equation}
then the tensor $\lambda^{-1}{\bf J}_\lambda$ is a Poisson one, and the
quantity $\lambda$ is a reducing multiplier (according to Chaplygin).

Thus, we obtain the following Hamiltonian vector field
$$
\bv_\lambda=(\lambda^{-1}{\bf J}_\lambda)\nabla H,
$$
for which $\div\bv_\lambda\ne 0$ ($\div\lambda\bv_\lambda=0$). The
examples of Hamiltonian fields with nonzero divergency were almost not
discussed earlier. One particular case of the Poisson tensor
$\lambda^{-1}{\bf J}_\lambda$ was found by J.\,Hermans in~\cite{hermans}.
Hermans used his own system of reduced variables which slightly differs
from ours.

To clarify the behavior of a ball on a surface of revolution, we will
discuss the cases, when this surface is a paraboloid, a sphere, a cone,
and a cylinder. These problems were investigated by Routh in~\cite{Raus}
for~$\Om=0$. Here we will expand the results for~$\Om\ne0$.

\paragraph{A paraboloid of revolution.}
Suppose the ball's center of mass is moving on the paraboloid of
revolution~$z=c(x^2+y^2)$. In equations~\eqref{eq1-11} we, therefore, put
\begin{equation}
\label{eq3-10}
f(\gamma_3)=-\frac{1}{2c\gamma_3}.
\end{equation}
The density of the invariant measure~\eqref{eq1-12} (up to an unessential
factor) can be written as
\begin{equation}
\label{eq3-11}
\rho=\frac{1}{\gamma_3^6}.
\end{equation}
In this case, the two-dimensional system~\eqref{eq1-16} reads
\begin{equation}
\label{eq3-12}
\frac{dK_1}{d\gamma_3}=\frac{1}{2c\gamma_3^3}\Bigl(
K_2-\frac{1+2cR\gamma_3}{2cR\gamma_3}\mu\Omega\Bigr),\quad
\frac{dK_2}{d\gamma_3}=-\frac{D}{\mu+D}\Bigl(2c\gamma_3K_1+
\frac{(1+2cR\gamma_3^3)}{2cR\gamma_3^2}\mu\Omega\Bigr).
\end{equation}
For the variable $K_2$ we obtain a {\em homogeneous} second-order linear
equation whose coefficients are homogeneous functions of $\gamma_3$ (at
$\Omega\ne 0$)
$$
K_2''-\frac{1}{\gamma_3}K_2'+\frac{D}{\mu+D}\frac{1}{\gamma_3^2}K_2=
\frac{D(2+cR\gamma_3)}{(\mu+D)cR\gamma_3^3}\mu\Omega.
$$
Its general solution may be represented as~$\gamma_3^\alpha$,
$\alpha=\const$
\begin{equation}
\label{eq3-13}
K_2=c_1\gamma_3^{1-\nu}-c_2\gamma_3^{1+\nu}+\mu\Omega\Bigl(
1+\frac{2D}{cR(\mu+4D)}\frac{1}{\gamma_3}\Bigr), \quad \nu^2=\frac{\mu}
{\mu+D}.
\end{equation}
For the variable $K_1$ from~\eqref{eq3-12} we, similarly, obtain
\begin{equation}
\label{eq3-14}
K_1=\frac{\mu+D}{2cD}\Bigl(-(1-\nu)\gamma_3^{-1-\nu}c_1+
(1+\nu)\gamma_3^{-1+\nu}c_2\Bigr)-\frac{\mu\Omega}{2c}
\Bigl(1-\frac{\mu}{2cR(3\mu+4D)}\frac{1}{\gamma_3^3}\Bigr).
\end{equation}
Solving for the constants~$c_1$, $c_2$ from~\eqref{eq3-13},
\eqref{eq3-14}, we obtain the integrals of the system~\eqref{eq1-16}.
These integrals are linear in to~$K_1$, $K_2$ and have form
\begin{equation}
\label{eq3-15}
\begin{aligned}
F_2&=\frac{D}{2\sqrt{\mu(\mu+D)}}\gamma_3^\nu\biggl(
2c\gamma_3K_1+\frac{\mu+D}{D\gamma_3}K_2+\mu\Omega\Bigl(
\gamma_3-\frac{(\mu+D)(1+\nu)}{D\gamma_3}-\frac{2+\nu}{2cR(2-\nu)
\gamma_3^2}\Bigr)\biggr)\\
F_3&=\frac{D}{2\sqrt{\mu(\mu+D)}}\gamma_3^{-\nu}\biggl(
2c\gamma_3K_1+\frac{\mu+D}{D\gamma_3}K_2+\mu\Omega\Bigl(
\gamma_3-\frac{(\mu+D)(1-\nu)}{D\gamma_3}-\frac{2-\nu}{2cR(2+\nu)
\gamma_3^2}\Bigr)\biggr)
\end{aligned}
\end{equation}
The product $F_2 F_3$ gives {\em an algebraic quadratic integral}.

For~$\Omega=0$ the equations also have the integral of energy
\begin{equation}
\label{eq3-16}
H=\frac{2c^2\gamma_3^2}{\mu}K_1^2+\frac12\frac{\mu+D}{\mu^2(1-\gamma_3^2)}
\bigl(2c\gamma_3^2K_1+K_2\bigr)^2+\frac12\frac{K_3^2}{\mu+D}.
\end{equation}

\begin{rem}
\label{rem2} Some forms of surfaces  (on which a ball is rolling) are
investigated in the paper~\cite{Bychkov} (details are given below). In
this paper it is shown that if a ball is rolling on a paraboloid of
revolution, then the system~(\ref{eq1-3}) is reduced to a particular class
of Fuchsian equations. Routh himself considered the case when not the
contact point, but the center of mass is moving on a paraboloid. We should
note again that in this case equations~(\ref{eq1-3}) have algebraic
integrals. The motion of a homogeneous ball on a surface of revolution was
also studied by F.\,Noether~\cite{neter}.
\end{rem}

\paragraph{An axisymmetric ellipsoid.}
Consider the motion of a dynamically symmetrical ball when its center of
mass is moving on a fixed ellipsoid ($\Om=0$). In this case
\begin{equation}
\label{eq3-29}
f(\gamma_3)=\frac{b_1}{\sqrt{b_1(1-\gamma_3^2)+b_3\gamma_3^2}},
\end{equation}
where $b_1,b_2$ are the squares of the ellipsoid principal semi-axes. The
density of the invariant measure~\eqref{eq1-12} (up to a constant factor)
is
\begin{equation}
\label{eq3-29a}
\rho=\bigl(b_1(1-\gamma_3^2)+b_3\gamma_3^2\bigr)^{-3}.
\end{equation}

In this case the variables $N_1,N_2,K_3$ are more convenient than the
variables~\eqref{eq1-14}. Here
\begin{equation}
\label{eq3-30}
N_1=(\bM,\bs\gamma),\quad N_2=\frac{\mu}{\mu+D}f(\gamma_3)\bigl(
\gamma_3(\bM,\bs\gamma)-M_3\bigr),
\end{equation}
which satisfy the system, similar to \eqref{eq1-15} for $\Omega=0$
\begin{equation}
\label{eq3-31}
\begin{gathered}
\dot N_1=-kK_3\frac{f'}{\gamma_3f^2}N_2,\quad \dot N_2=kK_3\frac{\mu}
{\mu+D}fN_1,\\
\dot K_3=-k\frac{(\mu+D)g}{\mu^2\gamma_3(1-\gamma_3^2)^2f^3}N_2
\bigl(\mu f(1-\gamma_3^2)N_1+(\mu+D)\gamma_3N_2\bigr),
\end{gathered}
\end{equation}
where $k=\frac{R\sqrt{1-\gamma_3^2}}{(\mu+D)g}$.\goodbreak

Using~\eqref{eq3-29}, we obtain two linear equations with independent
variable~$\gamma_3$
\begin{equation}
\label{eq3-32}
\frac{dN_1}{d\gamma_3}=-\frac{(b_1-b_3)}{b_1\sqrt{b_1(1-\gamma_3^2)+
b_3\gamma_3^2}}N_2,\quad \frac{dN_2}{d\gamma_3}=\frac{\mu b_1}
{(\mu+D)\sqrt{b_1(1-\gamma_3^2)+b_3\gamma_3^2}}N_1.
\end{equation}
It is easy to show that system~\eqref{eq3-32} has a quadratic integral
with constant coefficients
\begin{equation}
\label{eq3-33} F_2=b_1^2\frac{\mu}{\mu+D}N_1^2+(b_1-b_3)N_2^2.
\end{equation}
Below we will show that this integral may be generalized to the case of a
three-axial ellipsoid.

The system~\eqref{eq3-32} can be solved in terms of elementary functions.
Depending on the sign of the difference~$b_1-b_3$, the solution may be
written as:
\begin{itemize}
\item[1.] {\boldmath $b_1>b_3$, $a^2=\frac{b_1}{b_1-b_3}>1$.}
\begin{equation}
\label{eq3-34}
\begin{gathered}
N_1=c_1\sin\vfi(\gamma_3)+c_2\cos\vfi(\gamma_3),\quad N_2=
a\sqrt{\frac{\mu b_1}{\mu+D}}\bigl(-c_1\cos\vfi(\gamma_3)+
c_2\sin\vfi(\gamma_3)\bigr),\\
\vfi(\gamma_3)=\nu\arctg\frac{\gamma_3}{\sqrt{a^2-\gamma_3^2}},\qq
\nu=\sqrt{\frac{\mu}{\mu+D}}.
\end{gathered}
\end{equation}
\item[2.] {\boldmath $b_1<b_3$,} {\boldmath $a^2=\frac{b_1}{b_3-b_1}>0$.}
\begin{equation}
\label{eq3-35}
\begin{gathered}
N_1=c_1\tau^{-\nu}+c_2\tau^{\nu},\quad N_2=a\sqrt{\frac{\mu
b_1}{\mu+D}}\Bigl(-c_1\tau^{-\nu}
+c_2\tau^{\nu}\Bigr),\\
\tau(\gamma_3)=\gamma_3+\sqrt{a^2+\gamma_3^2},\qq
\nu=\sqrt{\frac{\mu}{\mu+D}}.
\end{gathered}
\end{equation}
\end{itemize}
Here $c_1, c_2=\const$. Solving for these constants, we can obtain linear
integrals of motion.

{\small
\paragraph{Historical comments.}
It is an interesting fact that neither Routh, nor his followers succeeded
in obtaining the simplest reduced equations (like~\eqref{eq3-32}) and
solving the problem of a ball rolling on an ellipsoid of revolution in
terms of elementary functions. To integrate the equations one must
appropriately choose the reduced variables such as~\eqref{eq3-30}.

}

\wfig<bb=0 0 50.7mm 30.9mm>[8]{ris4.eps}

\paragraph{A circular cone.}
In this case, due to the fact that the Gaussian map $\bs\gamma=
\frac{\nabla F}{|\nabla F|}$ is degenerate, one should use the components
of the vector~$\br$ (the position vector of the ball center) as the
positional variables in equations~\eqref{eq1-10}. For the cone~$\ta$
(Fig.~\ref{ris4.eps}) we have
\begin{equation}
\label{eq3-17}
\begin{gathered}
\gam_3=\cos\ta=\const, \q \gam_1=\frac{k^2}{\sqrt{1+k^2}} \frac{r_1}{r_3},
\q \gam_2=\frac{k^2}{\sqrt{1+k^2}} \frac{r_2}{r_3}, \\
r_3=k\sqrt{r_1^2+r_2^2},\qquad k=\tg\ta.
\end{gathered}
\end{equation}

\noindent The measure of equations~\eqref{eq1-3}, \eqref{eq1-10} in which
$\bs\gamma$ is expressed in terms of~$\br$ in accordance
with~\eqref{eq3-17}, can be written in explicit form: \eq[eq3-18]{
\rho=\frac{\left(\frac{k^2}{1-k^2}R_0+r_3\right)^3}{r_3^2}, }

\noindent
For the reduced system let us choose the variables
\begin{equation}
\label{eq3-19}
\begin{aligned}
\sg_1 & =\omega_3 +\frac{D\Omega}
{\sqrt{1+k^2}\sqrt{\mu+D}} \frac{r_3}{R},\\
\sg_2 & =\Bigl(r_3+\frac{k^2}{\sqrt{1+k^2}} R\Bigr)
\Bigl(\Bigl(\bM-\frac{k^2\mu^2}
{\mu+D}\bs\Omega\Bigr),\bs\gamma\Bigr).
\end{aligned}
\end{equation}
In terms of these variables we obtain the equations \eq*{
\frac{d\sg_1}{dr_3}=0, \q
\frac{d\sg_2}{dr_3}=\sqrt{1+k^2}\sg_1+\frac{\mu\Om}{\mu+D}\Bigl(\frac{r_3}{R}-k^2\Bigr).
} From these equations the following integrals can be easily obtained:
\begin{equation}
\label{eq3-20}
F_2=\sigma _1,\q
F_3=\sqrt{1+k^2}r_3\sigma_1-\sigma_2+\frac{\mu\Omega}{\mu+D}
\Bigl(\frac{r_3^2}{2R}-k^2r_3\Bigr).
\end{equation}

\paragraph{A circular cylinder.}
{\it The motion of a ball within a cylinder} is a well-known problem which
is usually used to illustrate some unrealistic conclusions, derived by
means of nonholonomic mechanics. It can be shown that a homogeneous ball,
moving within a vertical cylinder, on the average, is not rolling
downwards due to the gravity force. Nevertheless, this physical fact may
be observed while playing basketball, when the ball has almost hit the
basket, but then rapidly jumps out of it, suddenly lifting upwards. The
addition of a viscous friction to this nonholonomic system, which leads to
a vertical drift, is analyzed in~\cite{disskoleskinov}, where the explicit
solution of this problem has also been obtained.

\wfig<bb=0 0 23.7mm 37.1mm>{ris1.eps}

For a
cylinder,~$\bs\gam=\left(-\frac{r_1}{R_c},\,-\frac{r_2}{R_c},\,0\right)$,
where $R_c$ is the cylinder radius (Fig.~\ref{ris1.eps}), in terms of the
variables~$(\bM,\,\br)$ or~$(\bs\om,\,\br)$, the invariant measure's
density is constant. The kinetic energy looks like
\begin{multline*}
H=\frac12(\bM,\,\bs\om)=\frac{1}{2(\mu+D)} \left\{ (M_1^2+M_2^2)+
\frac{D}{\mu}(\bM,\,\bs\gam)^2+M_3^2 \right\} = \\
=\frac12(\mu(\bs\om,\bs\gam)^2 + (\mu+D)
(\om_3^2+(\om_1\gam_2-\om_2\gam_1)^2)).
\end{multline*}
The reduced system in terms of the variables~$\sg_1=\om_3$,
$\sg_2=(\bs\om,\,\bs\gam)$ looks like \eq[67]{ \sg_1'=\om_3'=0, \qq
\sg_2'=\frac{R\om_3-R_c\Om}{(R_c-R)R}. }

\noindent Thus, we have two integrals \eq[68z]{ \om_3=\const, \qq
(\bs\om,\bs\gam)-\frac{R\om_3-R_c\Om}{R_c-R} \frac{r_3}{R} = \const. } On
writing~$\wt{\bs\om}= (\om_1,\,\om_2,\,\frac{R\om_3-R_c\Om}{R_c-R})$, the
second integral takes the form \eq*{ (\wt{\bs\om},\br)=\const, } and the
kinetic energy becomes \eq*{
2H=\mu\left((\wt{\bs\om},\bs\gam)+\wt\om_3\frac{r_3}{R} \right)^2
+(\mu+mR^2){\wt\om}_3^2+(\mu+mR^2) \frac{\dot r_3^2}{R^2}. } Hence, the
variable~$r_3$ (responsible for the vertical displacement of the ball) can
be easily found to be \eq*{ r_3=-\frac{(\wt{\bs\om},\bs\gam)}{\wt\om_3}\pm
\sqrt{\vphantom{\frac{\mu}{\mu+mR^2}} \raise
-3pt\hbox{$\ds\smash{\frac{2H-(\mu+mR^2){\wt\om}^2_3}{\mu{\wt\om}^2_3}}$}}
\sin\Bigl({\wt\om}_3\sqrt{\frac{\mu}{\mu+mR^2}}(t-t_0)\Bigr), }

\wfig<bb=0 0 41.1mm 45.6mm>{ris2.eps}

\noindent where $t_0$ is a constant depending on the initial conditions.
It is clear that the average displacement of the ball equals zero even in
the presence of the gravity field (see formula~\eqref{eq-a7}).

Let us now consider two different variants of problem concerning the
rolling of a ball on a sphere. These problems were integrated and analyzed
by Routh~\cite{Raus}.

\paragraph{A ball on a rotating sphere.}
Consider a sphere of radius $R_s$ rotating around a certain axis with a
constant angular velocity~$\bs\Om$. Let $R$ be the ball radius,
$\ba=-R\bs\gam$, $\br=R_s\bs\gam$ (Fig.~\ref{ris2.eps}). The equations of
motion in the potential field with potential~$V(\bs\gam)$ may be
represented as \eqc[68]{
D_1\dot{\bs\om}=\frac{DR}{R_s+R}(\bs\om,\bs\gam)\bs\om\x\bs\gam+
\frac{R_sR}{R_s+R}
(\bs\Om,\bs\gam) (R\bs\om+R_s\bs\Om)\x\bs\gam+\bs\gam\x\pt{V}{\bs\gam}, \\
\dot{\bs\gam}=\frac{(R\bs\om+R_s\bs\Om)\x\bs\gam}{R_s+R} } where~$D=m
R^2$, $D_1=\mu+D$, and $\mu$ is the moment of inertia of the ball. Whether
the ball rolls outside or inside the sphere is determined by the sign
of~$R$. Using the angular momentum
vector~$\bM=D_1\bs\om-D\bs\gam(\bs\om,\bs\gam)$, the first equation
of~\eqref{68} may be written as
\begin{equation}
\label{69}
\dot \bM=\frac{R_sR}{R_s+R} \Bigl( R((\bs\om\x\bs\gam)(\bs\Om,\bs\gam)+
\bs\gam(\bs\Om,\bs\om\x\bs\gam))
-R_s(\bs\Om,\bs\gam)(\bs\gam\x\bs\Om) \Bigr) + \bs\gam\x\pt{V}{\bs\gam}.
\end{equation}
Using~$\wt{\bs\om}=\bs\om+\frac{R_s}{R}\bs\Om$ in~(\ref{68}), we get
\eq[70]{ \left\{
\begin{aligned}
D_1\dot{\wt{\bs\om}} & =\frac{DR}{R_s+R} (\wt{\bs\om},\bs\gam)
(\wt{\bs\om}\x\bs\gam)+\bs\gam\x\pt{V}{\bs\gam} \\
\dot{\bs\gam} & = \frac{R}{R_s+R}\wt{\bs\om}\x\bs\gam.
\end{aligned}
\right. } This coincides with the original system~\eqref{68}
for~$\bs\Om=0$. Thus, it will be enough to consider only the case
when~$\bs\Om=0$. In terms of the variables~$(\bM,\bs\gam)$
equations~(\ref{68}) for $\bs\Omega=0$ are written as \eq[71]{ \left\{
\begin{aligned}
\dot \bM & = \bs\gam\x\pt{V}{\bs\gam} \\
\dot{\bs\gam} & = \frac{R}{R_s+R}\bs\om\x\bs\gam.
\end{aligned}
\right. } They possess the integrals \eq[72]{
H=\frac12(\bM,\bs\om)+V(\bs\gam), \q F_1=(\bM,\bs\gam)=c, \q
F_2=\bs\gam^2=1. } The change of time~$t\to-\frac{R}{R_s+R}\tau$
transforms the second equation of~(\ref{71}) into an ordinary Poisson
equation~$\dot{\bs\gam}=\bs\gam\x\bs\om$, and the potential is multiplied
by some nonessential factor. Thus we set a system~(\ref{68}) for which we
know the cases when it is integrable. For example,
if~$V=\frac12(\bs\gam,{\bf C}\bs\gam)$, ${\bf C}=\diag(c_1,\,c_2,\,c_3)$,
one gets famous Neumann problem of the motion of a point on a sphere in
the quadratic potential (equations~(\ref{68}) incorporate even more
general situation, when~$(\bM,\bs\gamma)=c\ne0$, which corresponds to the
Clebsch case~\cite{BorisovMamaev2}).

The equations~\eqref{70}, \eqref{71}for the motion of a ball on a sphere
also manifests an analogy (discussed in~\cite{Neimark,Raus}) of problems,
concerning the rolling of a homogeneous ball on a sphere in the gravity
field, and the Lagrange case in the Euler\f Poisson equations (the motion
of a heavy dynamically symmetrical top). Indeed, for a potential of the
type~$V=V(\gam_3)$, the systems~\eqref{70}, \eqref{71}, because of the
axial symmetry, have ``Lagrangian integrals''~$M_3=\const$
or~$\om_3=\const$.

\paragraph{The rolling of a ball on a rotating sphere.}
Now suppose that the sphere on which a ball rolls rotates freely about its
center. The dynamical equations can be written as: \eqc*{ m\dot\bv =
\bN,\quad \mu\dot{\bs\om} = \ba\x \bN,\quad
\mu_s\dot{\bs\Omega}=-\br\x \bN, \\
\bv+\bs\om\x \ba  = \bs\Om\x \br. } Here~$\bs\om,\,\bs\Om,\,\mu,\,\mu_s$
are the angular velocities and the moments of inertia of the ball and the
sphere, respectively. Using the relations~$\br=R_s\bs\gam$,
$\ba=-R\bs\gam$, for the
quantities~$\wt{\bs\om}=\frac{R\bs\om+R_s\bs\Om}{R+R_s}$, $\bs\gam$, we
can write \eqc[73]{
\begin{gathered}
(1+D)\dot{\wt{\bs\om}}  = D\bs\gam\x(\dot{\bs\gam}\x\wt{\bs\om}),\quad
\dot {\bs\gamma}=\wt{\bs\omega}\x\bs\gamma, \\
D=\frac{mR^2}{\mu}+\frac{mR^2_s}{\mu_s}.
\end{gathered} \\
} By the change of time $dt\to\alpha\,dt$, $\alpha=\const$, this system
can be reduced to equations~\eqref{70} and, for this reason, is
integrable.

\section{Rolling of a ball on surfaces of the second order}

Consider the dynamics of a ball in greater detail for the case when its
center of mass is moving on a surface of the second order
\begin{equation}
\label{eq4-1}
\bigl(\br+R\bs\gamma,{\bf B}^{-1}(\br+R\bs\gamma)\bigr)=1,\quad
{\bf B}=\diag(b_1,b_2,b_3).
\end{equation}
In this case the equations of motion are identical in form
to~\eqref{eq1-8}.

It can be shown that these equations possess an invariant measure and a
quadratic integral of the form
\begin{equation}
\label{eq4-2}
\begin{gathered}
\rho=(\bs\gamma,{\bf B}\bs\gamma)^{-2},\\
F_2=\frac{(\bs\gamma\times \bM,{\bf B}^{-1}(\bs\gamma\times\bM))}
{(\bs\gamma,{\bf B}\bs\gamma)}.
\end{gathered}
\end{equation}
Here $\bf B$ is an arbitrary (nondegenerate) matrix.

{\small
\paragraph{Comment I.}
The invariant measure's density was found rather easily, after the authors
had obtained equations~\eqref{eq1-8} for rolling of a homogeneous ball on
an ellipsoid. These equations describe the {\em Jacobi nonholonomic
problem}. The problem's name stems from the following fact. When the
ball's radius is tending to zero it seems that we get the holonomic
classical problem concerning the geodesic lines on an ellipsoid (solved by
Jacobi in terms of elliptic functions). Apparently, one should get
convinced in the correctness of such a limiting transition, which,
however, does not prevent us from using the given terminology. One can
treat the integral~\eqref{eq4-2} as a generalization of the Ioachimstal
quadratic integral in the Jacobi problem. Originally, the authors found
this integral by numerical experiments, using the Poincar\'{e}
three-dimensional map in terms of the Andoyae\f Deprit variables
$(L,G,H,l,g,h)$. These variables for nonholonomic mechanics were
introduced in~\cite{Bormam-new} (see also earlier paper~\cite{BE}).

Fig.~\ref{map.eps} illustrates three-dimensional sections of a phase
flow in the phase space $(l,L/G,H,g)$ at a level of energy~$E=\const$. The
secant plane is $g=\pi/2$. It is seen that the levels~$F_2=\const$ make
the three-dimensional chaos ``foliated'' into two-dimensional chaotic
surfaces. The fact that the motion on the two-dimensional surfaces
$F_2=\const$ ensures that no additional integrals (necessary for
integrability of the problem) exist.

}

Let us consider various particular (may be degenerate) surfaces of the
second order for which an integral of the type~\eqref{eq4-2} exists.

\fig<width=15.8cm>{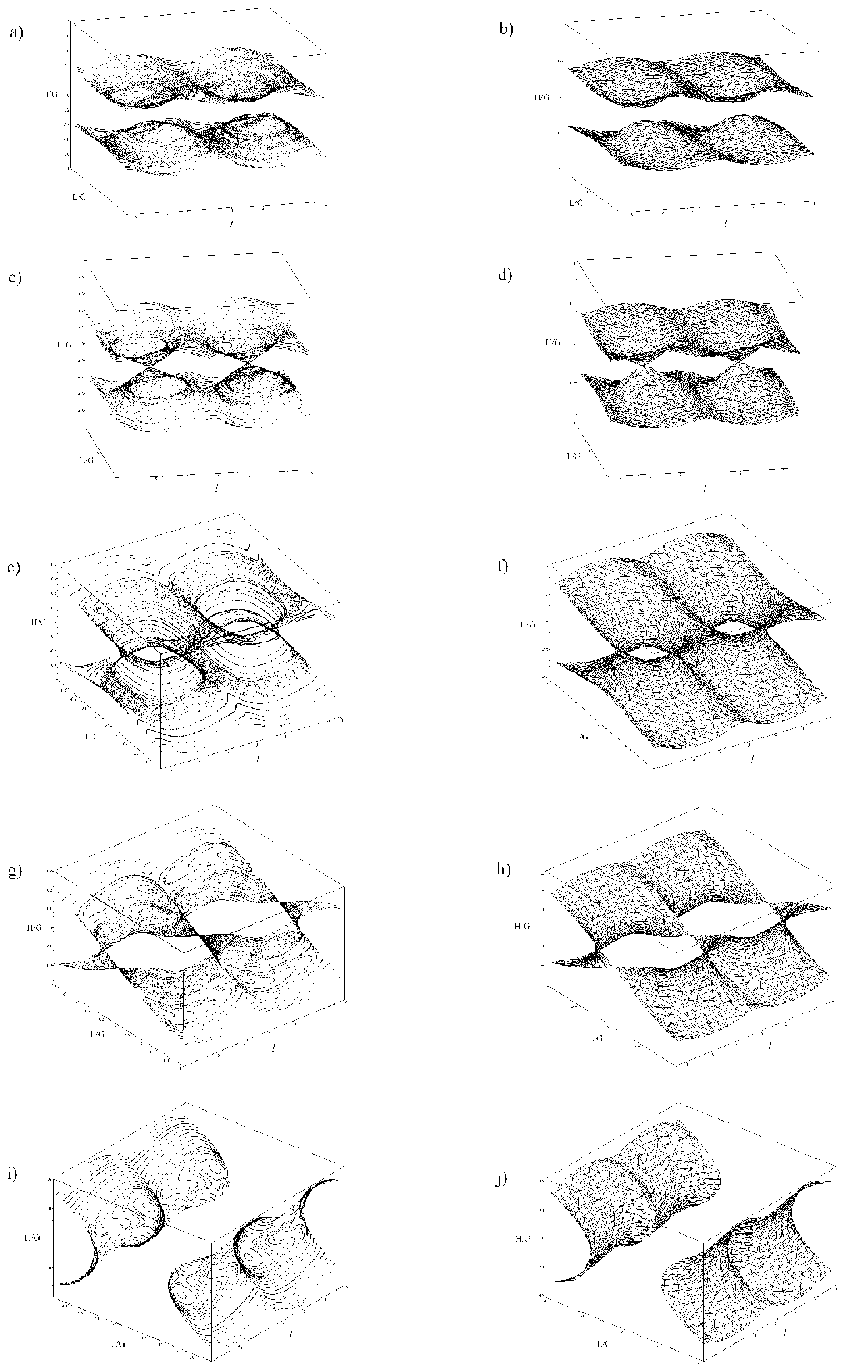} [\label{fig1}Examples of three-dimensional
mappings (on the left) and the corresponding level surfaces of the
integral $F_2$ (on the right). All the mappings are constructed at a level
of energy $E=1$ for ${\bf B}=\diag(1,\,4,\,9)$. To the frames on the left
correspond the values $F_2=1.7,\,2,\,4,\,8,\,10$ (from top to
bottom).\vspace{-16mm}]

\paragraph{An elliptic (hyperbolic) paraboloid.}
Let the ball's center of mass move on an elliptic paraboloid defined by
the equation
\begin{equation}
\label{eq4-3}
\frac{x^2}{b_1}+\frac{y^2}{b_2}=2z.
\end{equation}
Although in this case all the results may be obtained from the previous
ones by means of a passage to the limit, we will derive them ``of
scratch''.

The Gaussian map~\eqref{eq1-7} looks like:
\begin{equation}
\label{eq4-4}
r_1+R\gamma_1=-b_1\frac{\gamma_1}{\gamma_3},\quad r_2+R\gamma_2=
-b_2\frac{\gamma_2}{\gamma_3},\quad r_3+R\gamma_3=
\frac{b_1\gamma_1^2+b_2\gamma_2^2}{2\gamma_3^2},
\end{equation}
and the equations motion~\eqref{eq1-8} take the form
$$
\dot\bM=-\frac{D}{\mu+D}(\bM,\dot{\bs\gamma})\bs\gamma,\quad
\dot{\bs\gamma}=\frac{R\gamma_3}{\mu+D}\bs\gamma\times\bigl(
\bs\gamma\times{\bf B}^{i}(\bs\gamma\times\bM)\bigr),
$$
where ${\bf B}^{i}=\diag(b_1^{-1},b_2^{-1},0)$ is a degenerate
matrix.\goodbreak

The invariant measure density depends on~$\gamma_3$ only
\begin{equation}
\label{eq4-5}
\rho=\frac{1}{\gamma_3^4},
\end{equation}
and the quadratic integral~\eqref{eq4-2} can be written as
\begin{equation}
\label{eq4-6}
F_2=\frac{\bigl(\bs\gamma\times\bM,{\bf B}^{i}(\bs\gamma\times\bM)\bigr)}
{\gamma_3^2}.
\end{equation}

{\small
\paragraph{Comment II.}
The integrals~\eqref{eq4-2}, \eqref{eq4-6}, being quadratic with respect
to the velocities ($\bM$ or $\bs\omega$), depend on the positional
variables in a rather complex way. Maybe for this reason the classics (in
particular, Routh and F.\,Noether who obtained only particular results)
did not find these integrals. As it has been already noted, the
integrals~\eqref{eq4-2}, \eqref{eq4-6} were originally found through
numerical experiments. Their analytic form was obtained by means of the
following considerations.

As we have shown above, the problem concerning the rolling of a ball on a
paraboloid of revolution such that $b_1=b_2$ is integrable and has two
additional linear integrals. This integrals (in the absence of rotation
$\Om=0$) may be written as follows \eqa[eq4-7]{
I_1=\gam_3^{\sqrt{1-k}-1}(\sg_1-\sg_2\gam_3+\sg_2\gam_3\sqrt{1-k}),\\
I_2=\gam_3^{-\sqrt{1-k}-1}(\sg_1-\sg_2\gam_3-\sg_2\gam_3\sqrt{1-k}),\quad
k=\frac{D}{\mu+D} } where $\sg_1=\om_3$, $\sg_2=(\bs \omega,\,\bs\gam)$.
The product of these integrals is also an integral which is quadratic with
respect to $\sigma_i$, and a rational function of $\sg_1,\,\sg_2$ and
$\gam_3$ \eq[eq4-8]{
J=\frac{(\sg_2\gam_3-\sg_1)^2}{\gam_3^2}-\sg_2^2(1-k)=\frac{\sg_1^2}{\gam_3^2}-
2\sg_2\frac{\sg_1}{\gam_3}+k\sg_2^2. } Let us eliminate, using the
expression for energy~(\ref{eq1-18}), the term $k\sg_2^2$ from the
integral~(\ref{eq4-8}), and consider the integral \eq[eq4-9]{
F_2=J+2E=\frac{\om_3^2}{\gam_3^2}-2\sg_2\frac{\om_3}{\gam_3}+\bs\om^2. }
Substituting the expressions for $\sg_2$ into~(\ref{eq4-9}) and isolating
perfect squares, we obtain \eq[eq4-10]{
F_2=\frac{(\gam_2\om_3-\gam_3\om_2)^2+(\gam_3\om_1-\gam_1\om_3)^2}{\gam_3^2}.
} The integral $F_2$, written in such a form, is easily generalized to the
case of an arbitrary paraboloid ($b_1\ne b_2$) \eq[eq4-11]{
F_2=\frac{\frac{1}{b_1}(\gam_2\om_3-\gam_3\om_2)^2+\frac{1}{b_2}
(\gam_3\om_1-\gam_1\om_3)^2}{\gam_3^2}= \frac{(\bs\gam\x\bs\om,{\bf
B}^{i}(\bs\gam\x\bs\om))}{\gam_3^2},\quad {\bf
B}^{i}=\diag(b_1^{-1},b_2^{-1},0) } and any surface of the second
order~\eqref{eq4-2}. The integrals~\eqref{eq4-2}, \eqref{eq4-6} may be
used for stability analysis of stationary motions of a ball near the
points of intersection of the surface with the principal axes. The
symmetrical case of this problem was considered by Routh~\cite{Raus}.

}

\paragraph{Motion of a ball on an elliptic cone.}
Suppose that the ball's center of mass is moving on the surface of an
elliptic cone, defined by the equation \eqc[con1]{ (\br_c,\,{\bf
B}^{-1}\br_c)=0,\qq {\bf B}=\diag(b_1,b_2,-1), } where
$\br_c=\br+R\bs\gam$ are the coordinates of the center of mass, and
$b_1,\,b_2$ are positive quantities such that $\sqrt{b_1}$ and
$\sqrt{b_2}$ determine the slope of the generatrices with respect to the
coordinate axes. Given the coordinates of the center of mass, we can
calculate the normal to the surface $\bs\gam$ at this point as follows:
\eqc[con2]{ {\bs\gam}=\frac{{\bf B}^{-1}\br_c}{\sqrt{({\bf
B}^{-1}\br_c,\,{\bf B}^{-1}\br_c)}}. } In our case $\br$ (or $\br_c$) is
not uniquely defined by $\bs\gam$ (because $\bs\gam$ is constant on a
generatrix). Therefore, as phase variables, we will use not
$(\bM,\,\bs\gam)$ (as we did earlier), but $(\bM,\,\br_c)$. Upon
substitution of~(\ref{con2}) into the equations of motion~(\ref{eq1-3}),
we get the equations of motion in terms of the variables $\bM,\,\br_c$:
\eq[con3]{
\begin{cases}
\dot\br_c=\frac{R}{(\mu+D)\sqrt{({\bf B}^{-1}\br_c,\,{\bf B}^{-1}\br_c)}}
({\bM}\x{{\bf B}^{-1}\br_c}),\\
\dot{\bM}=\frac{DR}{(\mu+D)^2({\bf B}^{-1}\br_c,\,{\bf B}^{-1}\br_c)^{5/2}}
({\bf B}^{-1}(\bM\x{\bf B}^{-1}\br_c),\,{\bf B}^{-1}\br_c\x({\bf B}^{-1}\br_c\x\bM))
{\bf B}^{-1}\br_c.
\end{cases}
} Equations~(\ref{con3}) possess an energy integral
$$
H=\frac{1}{2(\mu+D)}(\bM^2+\frac{D(\bM,\,{\bf B}^{-1}\br_c)^2}
{\mu({\bf B}^{-1}\br_c,\,{\bf B}^{-1}\br_c)})
$$
and an invariant measure
$$
\rho=\sqrt{({\bf B}^{-1}\br_c,\,{\bf B}^{-1}\br_c)}.
$$

Let us now make one more change of variables and time \eqc*{ \by={\bf
B}^{-1}\br_c,\qq d\,\tau=\frac{R}{(\mu+D)\sqrt{({\bf B}^{-1}\br_c,\,{\bf
B}^{-1}\br_c)}}d\,t. } This results in \eq[con4]{
\begin{cases}
\by'={\bf B}^{-1}({\bM}\x\by),\\
{\bM}'=\frac{D}{(\mu+D)(\by,\,\by)^2}({\bf B}^{-1}(\bM\x\by),\,\by\x(\by\x\bM))\by
\end{cases}
} which possesses two ``natural'' integrals: the energy integral \eqc*{
H=\frac{1}{2(\mu+D)}({\bM}^2+\frac{D}{\mu}\frac{(\bM,\,\by)^2}{(\by,\,\by)}),
} and the geometrical integral \eqc*{ (\by,\,B\by)=0. } The latter defines
the surface in terms of the new variables along which the ball's center of
mass is moving. Moreover, equations~(\ref{con4}) possess invariant measure
with constant density. The generalization of the nontrivial
integral~(\ref{eq4-2}) to the case of equations~(\ref{con4}) looks like
\eq[con5]{ F_2=(({\bM}\x\by),\,{\bf B}^{-1}({\bM}\x\by)), } or, in terms
of the original physical variables, we set \eq*{ F_2=(({\bM}\x{\bf
B}^{-1}\br_c),\,{\bf B}^{-1}({\bM}\x{\bf B}^{-1}\br_c)). }

The question of existence of one more additional integral for
equations~\eqref{con4} remains open. Apparently, in the general case, when
$b_1\ne b_2$, it does not exist.

\section{Motion of a ball on a cylindrical surface}

Let us consider the rolling of a ball whose center of mass is moving on a
cylindrical surface. It can be shown that, in the absence of external
fields, this system may be integrated by quadratures; but when an external
force is directed along the cylinder generatrix, the equations are reduced
to a Hamiltonian system with one and a half degree of freedom .

\wfig<bb=0 0 25.9mm 37.8mm>{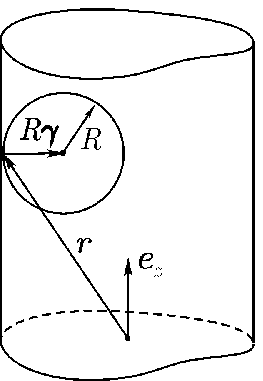}

Let us choose a fixed frame of reference with one axis~$(Oz)$ directed
along the cylinder generatrix (see Fig.~\ref{ris5.eps}). In this case, a
normal vector is expressed as
\begin{equation}
\label{eq-5-1}
\bs\gamma=(\gamma_1,\gamma_2,0),\quad \gamma_1^2+\gamma_2^2=1.
\end{equation}
Denote the projections of the normal to the center of mass and the
position vector of the center of mass of the ball onto the normal
cross-section by~$\wt{\br}=(r_1+R\gamma_1,r_2+R\gamma_2)$,
$\wt{\bs\gamma}= (\gamma_1,\gamma_2)$. For these projections we have
evident geometrical relations
$$
(\dot{\wt{\br}},\wt{\bs\gamma})=(\dot{\wt{\bs\gamma}},\wt{\bs\gamma})=0.
$$
Hence, we conclude that $\dot{\wt{\bs\gamma}}$ is parallel to
$\dot{\wt{\br}}$,
$$\dot{\wt{\bs\gamma}}=\lambda({\bs\gamma})\dot{\wt{\br}}.$$
The factor $\lambda({\bs\gamma})$ is completely determined by the geometry
of the cylinder cross-section and does not depend on angular velocity.

Using equations~\eqref{eq1-3}, we get the equations of motion for the ball
on the cylindrical surface with the assumption that the cylinder is
subject to a force (with potential~$U(z)$) directed along a cylinder's
generatrix ($z=\frac{r_3}{R}$):
\begin{equation}
\label{eq-a3}
\begin{gathered}
\dot\bM=\frac{M_3}{\mu+D}\lambda(\bs\gamma)\frac{D}{\mu+D}(\bM\times\bs\gamma,\be_z)\bs\gamma+
\pt{U}{z}\be_z\times\bs\gamma,\\
\dot{\bs\gamma}=\frac{M_3}{\mu+D}\lambda(\bs\gamma)\be_z\times\bs\gamma,\quad
\dot z=\frac{1}{\mu+D} (\bM\times\bs\gamma,\be_z).
\end{gathered}
\end{equation}

Here not only the energy is conserved but also the projection of the
(angular velocity) moment on the cylinder axis:
\begin{equation}
\label{eq-a4}
\begin{gathered}
H=\frac12(\bM,\bs\omega)+U(z),\\
F_2=M_3=(\mu+D)\omega_3=\const.
\end{gathered}
\end{equation}

Moreover, the system~\eqref{eq-a4} possesses an invariant measure with
density
\begin{equation}
\label{eq-a5} \rho({\bs\gamma})=\lambda^{-1}(\bs\gamma).
\end{equation}
It follows from~\eqref{eq-a3} that the equations for the
vector~$\bs\gamma$ get uncoupled. Let us use parametrization
$$
\gamma_1=\cos\vfi,\quad \gamma_2=\sin\vfi.
$$
For the angle $\vfi(t)$ we obtain the equation
\begin{equation}
\label{eq-a6}
\dot\vfi=\frac{M_3}{\mu+D}\lambda(\cos\vfi,\sin\vfi)=Q^{-1}(\vfi),
\end{equation}
where $Q(\vfi)$ is, in the general case, a $2\pi$-periodic function
of~$\vfi$, determined by the shape of a cylinder cross-section.

In the remaining equations of the system \eqref{eq-a3} we put
$$
K_1=M_1\gamma_1+M_2\gamma_1,\quad K_2=M_1\gamma_2-M_2\gamma_1,
$$
and replace the time (as an independent variable) by the angle~$\vfi$, and
thereby obtain the nonautonomous system with~$2\pi$-periodic coefficients
\begin{equation}
\label{eq-a7}
\begin{gathered}
\frac{dK_1}{d\vfi}=-\frac{\mu}{\mu+D}K_2,\quad \frac{dK_2}{d\vfi}=
K_1-Q(\vfi)U'(z),\\
\frac{dz}{d\vfi}=\frac{Q(\vfi)}{\mu+D}K_2.
\end{gathered}
\end{equation}
This system has an integral of energy
\begin{equation}
\label{eq-a8}
\wt{H}= \frac12\Bigl(\frac{K_1^2}{\mu}+\frac{K_2^2}{\mu+D}\Bigr)
+U(z).
\end{equation}
In the case of the gravity field, $U(z)=mgz$ and equation~\eqref{eq-a7}
are integrated by quadratures:
\begin{equation}
\label{eq-a9}
\begin{aligned}
K_1(\vfi)&=-\nu mg\intl_{\vfi_0}^{\vfi}\sin\nu(\tau-\vfi)Q(\tau)d\tau+
\nu A\cos\nu\vfi+\nu B\sin\nu\vfi,\\
K_2(\vfi)&=-mg\intl_{\vfi_0}^{\vfi}\cos\nu(\tau-\vfi)Q(\tau)d\tau+
A\sin\nu\vfi-B\cos\nu\vfi,
\end{aligned}
\end{equation}
where $A,B$ are constants and $\nu^2=\frac{\mu}{\mu+D}$.\goodbreak

Let us show that the integrals in~\eqref{eq-a9} are bounded functions. We
expand the function $Q(\tau)$ in a Fourier series \eq[eq-a9a]{
Q(\tau)=\sum\limits_{n\in\mZ}Q_ne^{in\tau}. } The integrals in the
expressions for $K_1(\varphi)$ and $K_2(\varphi)$ in~\eqref{eq-a9} may be
considered as the real and imaginary parts of the integral \eq[eq-a9b]{
\int e^{i\nu\tau}Q(\tau)\,d\tau=\int\sum\limits_n
Q_ne^{i(n+\nu)\tau}\,d\tau. } Using the well known theorems of the Fourier
analysis and the fact that $n+\nu\ne0$ (as far as $0<\nu<1$) we put the
integral under the sum sign and integrate the series term-by-term:
\eq[eq-a9c]{
\int\sum\limits_nQ_ne^{i(n+\nu)\tau}\,d\tau=\sum\limits_n\frac{Q_n}{i(n+\nu)}e^{i(n+\nu)\tau}.
} It is evident that the series obtained converges to a certain
quasiperiodic function, hence, $K_1(\varphi)$ and $K_2(\varphi)$ are
bounded. This fact and also the conservation of energy in the reduced
system~\eqref{eq-a8} ensure that $z(\vfi)$ is bounded.

{\it Thus, when the ball is rolling on an absolutely rough cylindrical
surface of an arbitrary cross-section in the gravity field, the vertical
secular drift is not observed.}

The time dependence of the angle (and, hence, of all the other functions)
is described by~\eqref{eq-a6}.

\paragraph{An elliptic (hyperbolic) cylinder.}
Let us consider in greater detail a particular case, i.\,e., the ball's
center of mass is moving on an elliptic cylinder with cross-section is
defined by the equation
\begin{equation}
\label{eq-**1}
\frac{x^2}{b_1}+\frac{y^2}{b_2}=1.
\end{equation}
We have
$$
r_1+R\gamma_1=\frac{b_1\gamma_1}{\sqrt{b_1\gamma_1^2+b_2\gamma_2^2}},\quad
r_2+R\gamma_2=\frac{b_2\gamma_2}{\sqrt{b_1\gamma_1^2+b_2\gamma_2^2}},\quad
r_3=Rz,
$$
therefore,
$$
\begin{gathered}
\lambda(\bs\gamma)=\frac{R(\bs\gamma,{\bf B}\bs\gamma)^{3/2}}{b_1b_2},\quad
{\bf B}=\diag(b_1,b_2,0),\\
Q^{-1}(\vfi)=\frac{M_3R}{(\mu+D)b_1b_2}(b_1\cos^2\vfi+b_2\sin^2\vfi)^{3/2}.
\end{gathered}
$$

We should mention an important distinction existing between an elliptic
and a circular cylinder (see above): in the case of an elliptic cylinder
the dependency of dynamical variables~$K_1,K_2,z$ is defined by two
frequencies~$\omega_1=1$, $\omega_2=\nu$, instead of a single frequency,
as it happens in case of a circular cylinder. Thus, the integrals
in~\eqref{eq-a9} contain quasiperiodic functions; the integrals have very
complicated nature, their analytical properties are thoroughly discussed
in~\cite{KozlovMKA}. Graphs $z(\varphi)$ for various initial values of
$K_1$ and $K_2$ are shown in Fig.~\ref{picuhod}. The main result is that
whatever the ratio of the frequencies, the quantities $K_1$ and $K_2$,
and, therefore, the displacement~$z$ execute bounded, quasiperiodic
oscillations. It is the main result of this construction.

The authors express are deeply thankful to V.\,V.\,Kozlov for useful
remarks and discussions.

\fig<width=110mm>{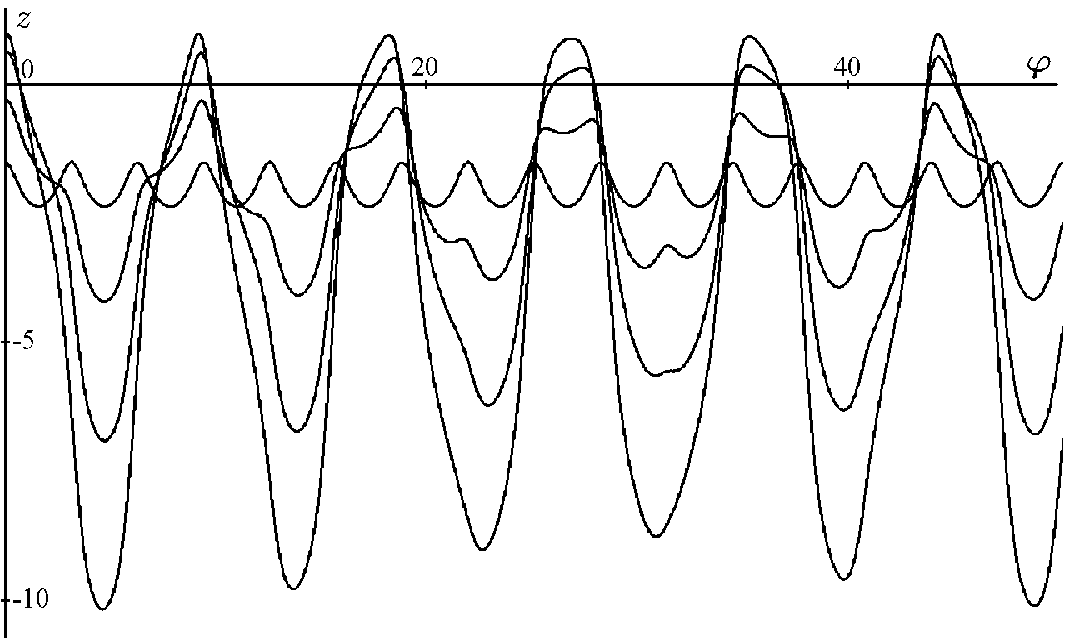}[\label{picuhod} The $\varphi$-dependence of
the vertical coordinate of the point of contact $z$ for various initial
values $K_1,\,K_2,\,z$. For this figure the other parameters are:
$E=1,\,\mu=1,\,D=1(\nu=2^{-1/2}),\,b_1=1,\,b_2=2,\,R=1$.]

\pagebreak


\begin{table}[!p]
\begin{center}
\unitlength=1mm
\begin{picture}(0,0)
\put(19.5,129.5){\cfig{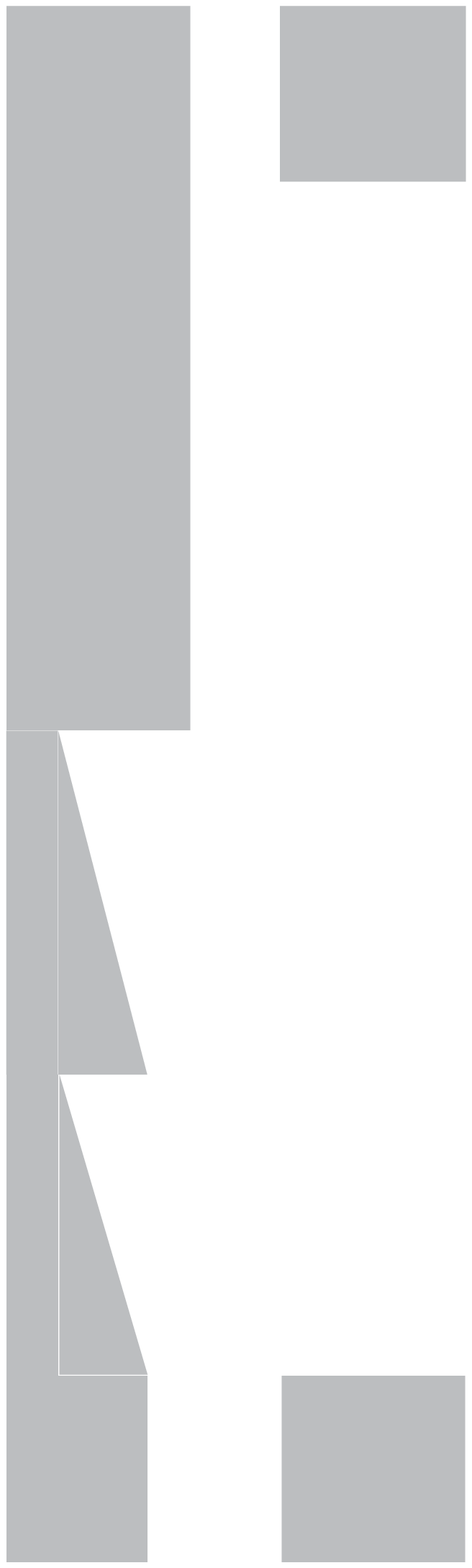}}
\end{picture}
\resizebox{!}{\textheight}{%
\rotatebox{90}{\parbox{27.5cm}{%
\begin{center}
\mbox{}\hfill{\small Table 1. The rolling of a ball on a surface}\\\medskip
\footnotesize\tabcolsep=1pt
\begin{tabular}{|c|c|c|c|c|c|c|c|c|c|}
\hline
& & \multicolumn{2}{|c|}{surface of the second order}
 & \multicolumn{6}{|c|}{surface of revolution}\\
 \hhline{|~|~|--|------|}
\parbox{22mm}{\centering surface type}
  & \parbox{30mm}{\centering \hspace{0pt}cylindrical surface}
  & \parbox{40mm}{\centering ellipsoid, hyperboloid,\\ paraboloid}
  & \parbox{40mm}{\centering cone of the second order}
  & \parbox{25mm}{\centering arbitrary\\ surface}
  & \parbox{22mm}{\centering ellipsoid,\\ hyperboloid}
  & \multicolumn{3}{|c|}{\parbox{40mm}{\centering paraboloid, cone, cylinder}}
  &   sphere\\
  \hline
  \multicolumn{10}{|c|}{}\\[-3.5mm]
  \hline
  \parbox{23mm}{\hspace{0pt}\centering  measure}& measure exists &
  \parbox{40mm}{\centering $\rho = (\bs\gamma,{\bf B}\bs\gamma)^{-2}$} &
  \parbox{35mm}{\centering $\rho=\sqrt{({\bf B}^{-1}\br_c,\,{\bf B}^{-1}\br_c)}$}&
  \multicolumn{6}{|c|}{\parbox{70mm}{\centering  $\rho = (f(\gamma_3))^3\Bigl(f(\gamma_3)-
  \frac{1-\gamma_3^2}{\gamma_3}f'(\gamma_3)\Bigr)$}}\\
  \hline
  \parbox{23mm}{\hspace{0pt}\centering additional integrals}
  &  \parbox{28mm}{\hspace{0pt}\centering system is integrable by quadratures}
  &  \parbox{48mm}{\centering$ \frac{(\bs\gamma{\times}\bM,{\bf B}^{-1}
  (\bs\gamma{\times}\bM))}{(\bs\gamma,{\bf B}\bs\gamma)}=\const$\\ (one integral)}
  &\parbox{55mm}{\centering$((\bM\x{\bf B}^{-1}\br_c),{\bf B}^{-1}(\bM\x{\bf B}^{-1}\br_c))=\const$\\
  (one integral)}
  &  \parbox{25mm}{\hspace{0pt}\centering
  two linear integrals, defined by the system of linear equations} &
  \multicolumn{5}{|c|}{\parbox{80mm}{\hspace{0pt}\centering
  there exist two linear integrals that can be
  expressed in terms of elementary functions}} \\
  \hline
  \parbox{23mm}{\hspace{0pt}\centering Hamiltonianity} & \multicolumn{3}{|c|}
  {\parbox{70mm}{\centering nothing is known about
  the Hamiltonianity of these systems}} & \multicolumn{6}{|c|}
  {\parbox{90mm}{\centering
  upon change of time (prescribed by the reducing multiplier)
  the reduced system becomes Hamiltonian}}\\
  \hline
  \parbox{23mm}{\hspace{0pt}\centering authors}&  \multicolumn{3}{|c|}{A.V.Borisov, I.S.Mamaev, A.A.Kilin (2001)}
  &  E.Routh (1884) &  \parbox{20mm}{\centering A.V.Borisov, I.S.Mamaev, A.A.Kilin (2001)} &
  \multicolumn{4}{|c|}{E.Routh (1884)}  \\
  \hline
  \parbox{22mm}{\hspace{0pt}\centering generalizations\\ and remarks} &
  \parbox{30mm}{\centering
  integrable addition of the gravity field
  along a cylinder generatrix is possible} & \multicolumn{2}{|c|}{}&
  \multicolumn{5}{|c|}{\parbox{85mm}{\centering
  A.V.Borisov, I.S.Mamaev, and A.A.Kilin have shown
  the integrability in terms of elementary functions
  of the case of the ball rolling on the ellipsoid of revolution;
  have found an invariant measure for an arbitrary surface of revolution;
  for a paraboloid, a cone, and a cylinder they have shown the integrability of
  the case, when the ball is rolling on a surface rotating around its
  axis of symmetry}}
  &  \parbox{28mm}{\centering
  the problems of the rolling of a ball on an
  unconstrained and rotating sphere are also solved.
  System also allows integrable additions of potentials.}  \\
  \hline
\end{tabular}
\end{center}
Remark. The cases when the tensor invariants exist are indicated by gray
color in the table. The partial filling corresponds to the uncomplete set
of integrals. }}}
\end{center}
\end{table}

\end{document}